\documentclass[cits]{PoS}

\usepackage{epsfig}
\usepackage{graphicx}

\def\slashchar#1{\setbox0=\hbox{$#1$}
   \dimen0=\wd0 \setbox1=\hbox{/} \dimen1=\wd1
   \ifdim\dimen0>\dimen1 \rlap{\hbox to \dimen0{\hfil/\hfil}} #1
   \else  \rlap{\hbox to \dimen1{\hfil$#1$\hfil}} / \fi}

\title{Transversity relations, chiral and holographic models, and 
  pion wave functions from lattice QCD}

\ShortTitle{Pion wave function}

\author{\speaker{Enrique Ruiz Arriola}
\thanks{Supported by the
  Spanish DGI and FEDER funds with grant FIS2008-01143/FIS, Junta de
  Andaluc{\'\i}a grant FQM225-05, and EU Integrated Infrastructure
  Initiative Hadron Physics Project contract RII3-CT-2004-506078.}\
\\ Departamento de F\'{\i}sica At\'omica, Molecular y
Nuclear, Universidad de Granada, \\ E-18071 Granada, Spain\\ E-mail:
\email{earriola@ugr.es}}

\author{Wojciech Broniowski
         \thanks{Supported by Polish Ministry of Science
                 and Higher Education, grants N~N202~263438 and N~N202~249235}\\
        The H. Niewodnicza\'nski Institute of Nuclear Physics, Polish Academy of Sciences, PL-31342~Krak\'ow, Poland\\
        Institute of Physics, Jan Kochanowski University, PL-25406~Kielce, Poland\\
        E-mail: \email{Wojciech.Broniowski@ifj.edu.pl}}

\abstract{We analyze the equal-time Bethe-Salpeter quark wave functions
  of the pion in various models. We discuss how the quenched lattice QCD results with
  delocalized pion interpolators can be identified with
  the coarse grained wave functions, typical of low-energy effective
  models. Actually, we find that one-loop chiral quark models predict
  that pseudoscalar and tensor wave functions have the same shape,
  while the axial component is more extended. These
  facts are accurately confirmed by the lattice. We also show how the
  transversity information, relevant for the light-cone physics, can be
  straightforwardly obtained from the equal-time rest-frame lattice 
  calculations. This
  remarkable relation provides a way to extract, for instance, 
  the equal-time holographic wave functions and compare them, quite favorably, to the
  lattice calculations.}

\FullConference{Light Cone 2010 - LC2010\\
		June 14-18, 2010\\
		Valencia, Spain}

\begin{document}

\section{Introduction}

Hadronic wave functions encode important information on bound states
in strong-interaction physics; in particular, they provide the
amplitude for a composite hadron to have quarks in a given momentum
state or, equivalently, at a certain space-time distance.  Heavy
quarks obey non-relativistic quantum mechanics and conserve particle
number. While our understanding and intuition is based on
wave functions, as a matter of principle the wave functions cannot be
directly measured experimentally.  One must instead resort to form
factors, decay widths, or momentum distributions. Moreover, for light
quark systems particle creation may occur, demanding a field-theoretic
framework where further complications arise. Relativistic invariance
requires that one uses the conventional Bethe-Salpeter (BS) amplitudes
with a fixed number of the quark field operators, a reminiscent of the
approximated parton picture point of view, emphasized by the
light-cone approaches~\cite{Brodsky:1997de}. Color gauge invariance
requires additional inclusion of the link
operators~\cite{Suura:1977cw}.

For the pion, the spontaneously broken chiral symmetry is a basic
dynamical ingredient in the determination of its nonperturbative quark
structure. It appears via the pertinent axial Ward-Takahashi
identities~\cite{Pagels:1979hd}. These important constraints are
implemented in relativistic field-theoretic chiral quark models, such
as the Nambu--Jona-Lasinio (NJL) model (for a review see,
e.g.,~\cite{RuizArriola:2002wr}). The regularization, introducing the
physical cut-off, needs to be carefully handled not to spoil the
relativistic, gauge, and chiral symmetries.

On the other hand, lattice QCD solves the bound state problem in a
fundamental way.  It is thus possible to make a first-principle
nonperturbative determination of the wave functions, but at the expense
of breaking the continuum symmetries, such as the Lorentz invariance
and, quite often, chiral symmetry, due to the finite lattice
spacing. The axial Ward-Takahashi identities can be exactly
implemented on the discrete Euclidean lattice as shown by Ginsparg and
Wilson~\cite{Ginsparg:1981bj} (see Ref.~\cite{Gattringer:2000js} for a
recent practical implementation), enabling realistically small pion
masses.

In the present contribution we show the analysis of the pion wave
functions from the quenched lattice QCD~\cite{Broniowski:2009dt} and
make the comparison to various hadronic models. In spite of the very
dissimilar appearance and nature of these approaches, we will provide
the conditions under which this comparison may be undertaken. We also
address in more detail the light-cone issues with the help of the
transversity relations.

\section{Bethe-Salpeter Amplitudes }

The BS vertex or wave function of the pion is given by
\begin{eqnarray}
\chi_p (k)  =- i \int d^4 x e^{-i k \cdot x}\langle 0 | T \left\{ q(x) \bar q(0) \right\} | \pi_a (p) \rangle ,
\end{eqnarray}
where $q(x)$ are spinor field operators carrying flavor and color,
and $| \pi_a (p) \rangle$ is the pion state with the Cartesian isospin
index $a$ and the on-shell four-momentum $p$, $p^2=m_\pi^2$. While chiral
quark model calculations are naturally formulated in the momentum space,
the basic objects in Euclidean lattice calculations are the
point-to-point correlation functions. These quantities are gauge and
renormalization-group invariant at all Euclidean times, which
basically correspond to off-shell processes. At large Euclidean times
only the on-shell states contribute to the correlation functions, as well as the
off-shell violations of the gauge invariance disappear. Inverting the
Fourier transformation we get (in the isospin limit of $m_u=m_d$)
\begin{eqnarray}
\langle 0 | T \left\{ q(x) \bar q(0) \right\} | \pi_a (q) \rangle =
\frac{\tau_a}{f_\pi} \gamma_5  [\Psi_P +
  \slashchar{q} \Psi_A &+& i \sigma^{\mu \nu} q_\mu x_\nu \Psi_T] ,
\label{eq:coor-x}
\end{eqnarray} 
where the wave functions $\Psi_{a}$, $a=P,A,T$, depend on the
Lorentz-invariant variables $x^2$, $x \cdot q $, and $q^2=m_\pi^2
$. The quantities $\Psi_{a}$ are vertex
functions in the BS equation, and as such are finite and undergo
$x$-independent {\em multiplicative} renormalization. Thus, the ratios
$\Psi_a(x)/\Psi_a(0)$ become cut-off independent, as the cut-off is
removed, which on the lattice means $a \to 0$.

The definition of Eq.~(\ref{eq:coor-x}) is
completely satisfactory for chiral quark models. In QCD, however, it
is only gauge-invariant in the fixed-point Fock-Schwinger gauge,
$x^\mu A_\mu(x)=0$, where the standard derivatives, $\partial^\mu$, and
the covariant derivatives, $D^\mu = \partial^\mu + i g A^\mu$,
coincide. On the lattice the gauge fixing has the problem of the Gribov
copies, as there is no complete gauge fixing. On the other hand, Elizur's theorem
prevents non-vanishing vacuum expectation values of gauge variant
operators in the physical Fock space. 

Non-gauge invariant operators can be made gauge invariant by joining
them with a link operator, however, as a result the path-dependence
sets in.  Furthermore, gluons carry momentum in the pion and different
gauge-invariant definitions yield different results (see
Ref.~\cite{Negele:2000uk} for a discussion on various
possibilities). For definiteness, we choose a straight-line path
and undertake a {\em smearing} procedure. This delocalization improves
the signal-to-noise ratio for the measured hadron correlators, as the
interpolating operators have a larger overlap with the desired
state. Local operators, in contrast, do not take into account the
spatial extension of the hadron. The usefulness of the smearing
process lies also in the fact that the overall thickness of the flux
tube in the probe is controlled by the number of the smearing
steps. In addition, the method is computationally simple. The
resulting {\em fat link} also reduces the high-energy fluctuations and
the path dependence, such that we deal with a coarse-grained wave
function. The procedure naturally finds its counterpart in the
low-energy effective chiral quark models. In a previous
work~\cite{Broniowski:2009dt} the quenched lattice calculations of the
pion have been worked out along these lines.  The quenched 
approximation contains all the leading-$N_c$, and hence the $\bar{q}
q$, Fock state components. Thus we expect that quenched calculations
describe the large-$N_c$ motivated models~\footnote{The quenched
  lattice calculations also contain a piece subleading in $N_c$, which
  is actually suppressed for heavy quarks;  pion
  loops are $1/N_c$-suppressed, although not all of the
  $1/N_c$-contributions originate from the pion
  loops~\cite{Cohen:1992kk}. }.

\section{Transversity relations}

The relativistically invariant BS amplitude has the representation
\begin{eqnarray}
\langle 0 | T \left\{ q(x) \bar q(0) \right\} | \pi_a (q) \rangle &=& 
 i \gamma_5 \tau_a  \int_0^1 \, d\alpha e^{-i (2\alpha-1) q \cdot x } \nonumber \\  &\times&   
 [ - \widetilde{\Psi}_P(\alpha, x^2) + \slashchar{q} \widetilde{\Psi}_A (\alpha, x^2) - 2i \sigma^{\mu \nu} q_\mu x_\nu \widetilde{\Psi}_T (\alpha,  x^2)], 
  \label{eq:coor-x-2}
\end{eqnarray}
where $\alpha$ is  the Feynman parameter. As a matter of principle,
all scalars such as $\Psi_{a}$ in Eq.~(\ref{eq:coor-x}) depend on the
kinematic variables $x^2$, $ x \cdot q $, and $q^2$, thus we are free
to choose any form of kinematics.  In the {\em rest-frame} kinematics
we have $x_0=0$ and $(q_0,{\bf q})=(m_\pi, 0) $, whence $x^2 = - {\bf
  x}^2$, $ x \cdot q =0$, and $q^2 = m_\pi^2 $. On the other hand, in
the infinite-momentum-frame kinematics $(q_0,{\bf
  q})=(\sqrt{m_\pi^2+p_z^2}, p_z)$, with $p_z \to \infty$.  Thus on
the light-cone surface, $x_0 = z$, one has $x^2=-{\bf x}_T^2$, and $x
\cdot q \to 0$.

The Lorentz invariance allows us to relate the rest-frame calculation
to the transverse-coordinate dependence in the light-cone wave
functions.  Simply, by comparing Eqs.~(\ref{eq:coor-x}) and
(\ref{eq:coor-x-2}) we find
\begin{eqnarray}
\Psi_a(x^2 , x.q) = \int_0^1 d\alpha \widetilde{\Psi}_a ( \alpha,x^2 )e^{- i q.x (2\alpha-1)}.  
\end{eqnarray}
We may identify the Feynman parameter $\alpha$ with the Bjorken
$x$-variable, $x_{\rm Bj} \equiv \alpha$.  Then, for the chosen
kinematics $q^+ x^-=q \cdot x=0$, we have $\Psi^{\rm ET}_a(-r^2,0) =
\int_0^1 dx_{\rm Bj} \Psi^{\rm LC}_a (x_{\rm Bj}, -r^2 )$, where ET
and LC denote the equal-time, and light-cone wave functions,
respectively.  In the argument of $\Psi^{\rm ET}$ one takes $r^2={\bf
  x}^2$, the distance squared, while in the argument of $\Psi^{\rm
  LC}$ we need to use $r^2={\bf x}_T^2$, the {\em transverse} distance
squared. Therefore
\begin{eqnarray}
\Psi^{\rm ET}_a(-r^2,0) = \int_0^1 dx_{\rm Bj} \left . \Psi^{\rm LC}_a (x_{\rm Bj}, -{\bf x}_T^2 ) \right |_{x_T=r}.
\label{eq:transv}
\end{eqnarray}
That way the connection between the ET and LC wave functions has been
established
\footnote{Note that we keep only the
  $\bar q q$ components of the BS amplitude, thus we do
  not account for the possible emergence of the higher Fock-state
  components, pertinent to the dynamical nature of the boost.}.

Although generating the autonomous connection between LC and ET wave
functions seems at first glance hopeless~\cite{Miller:2009fc}, a
similar transversity relation has recently been deduced for scalar
particles.
Another transversity
property for the Generalized Parton Distributions (GPDs) was suggested
for the nucleon~\cite{Miller:2007uy} and the
pion~\cite{Broniowski:2007si}, allowing a frame-independent definition
of probability (unlike the more conventional Breit-frame definition).
In the case of the pion it reads
\begin{eqnarray}
P^{\rm LC}({\bf b})=\int_0^1 dx \, q(x,{\bf b}) = \int \frac{d^2
  q_\perp}{(2\pi)^2} e^{i  {\bf q}_\perp \cdot {\bf b}} F_V (-q_\perp^2) = P^{\rm
 ET}(r)|_{r=b \equiv |{\bf b}|}
\end{eqnarray}
where $q(x,{\bf b})$ is the off-forward diagonal GPD and $F_V(t)$ is
the pion form factor. In the phenomenologically successful Vector
Meson Dominance (VMD), where $F_V(t)=M_V^2/(M_V^2-t)$, one gets
\begin{eqnarray}
P^{\rm ET}(r)  = M_V^2 K_0( r M_V)/(2 \pi) \sim e^{-M_V r}/r^\frac12.
\label{eq:prob-inv}
\end{eqnarray}
Besides these relations, it would also be useful to verify the ET-LC
transversity connection directly on the lattice. While there exist
transverse lattice calculations~\cite{Dalley:2002nj} (see
\cite{Burkardt:2001jg} for a review), their focus is placed on the
Distribution Amplitude, $\Psi(\alpha, 0) = \varphi(\alpha)$, leaving
out the $x_\perp$ dependence. 

\section{Chiral Quark Models vs Lattice}

\begin{figure}
\begin{center}
\epsfig{figure=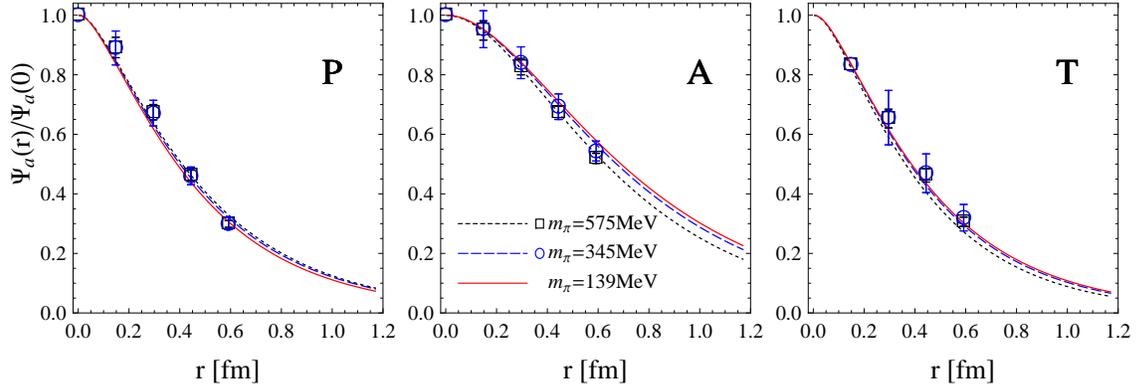,width=\textwidth}
\end{center}
\vspace{-6mm}
\caption{The components of the rest-frame equal-time pion wave function,
  normalized to unity at the origin, $\Psi_{a}({r})/\Psi_{a}({0})$,
  for $S$, $A$, and $T$ channels, evaluated in the NJL model at
  $M=g_{\pi qq} f_\pi=300$~MeV, and compared to the quenched lattice
  data~\cite{Broniowski:2009dt}. The displayed points of the lattice
  data are at $m_\pi=345$ and $575$~MeV, while the NJL model
  calculation includes also the case of the physical pion mass.}
\label{fig:njl-wf}
\end{figure}

We evaluate the correlation function of Eq.~(\ref{eq:coor-x}) in a
chiral quark model (for a review see,
e.g.,~\cite{RuizArriola:2002wr}).
Disregarding for the moment the regularization, 
an instructive way to determine the pion wave function in
a chiral quark model is by exploiting the axial Ward-Takahashi
identity. It relates the quark propagator, $S(p)$, and the vertex function
corresponding to the axial current, $J^{\mu,a}_A(x) =\frac12 \bar q
(x)\gamma^\mu \gamma_5 \tau_a q (x)$, with the irreducible vertex
$\Gamma_A^{\mu,a} (p+q,p)$:
\begin{eqnarray}
S(p+q)^{-1}  \gamma_5 \frac12 \tau_a  + \gamma_5 \frac12 \tau_a  S(p)^{-1}  
 = q_\mu  \Gamma_A^{\mu,a} (p+q,p) . 
\label{eq:axial0} 
\end{eqnarray}
In the NJL model, the spontaneous breaking of the chiral symmetry
generates a constituent quark mass, $M$, given by the so-called {\it
  gap equation}. As the result, $S(p)= 1/(\slashchar{p}-M)$, such that
\begin{eqnarray}
\Gamma_A^{\mu,a} (p+q,p) = \frac{\tau^a}2 \gamma_5 \left[\gamma^\mu -
  {q^\mu \over q^2} \frac{2 M}{f_\pi} \right] 
\end{eqnarray}
The pole at $q^2=0$ indicates the Goldstone boson nature of the pion.
The pion wave function is extracted from the pion pole as an unamputated
vertex function,
\begin{eqnarray} 
\chi_q^a (k) = { i \over \slashchar{k}+
\slashchar{q} - M } \left(\frac{M}{f_\pi} \gamma_5 \tau_a \right) { i
\over \slashchar{k} - M} \, , 
\end{eqnarray} 
where the Goldberger-Treiman relation at the quark level, $g_{\pi qq}=
M/f_\pi$, can be read off. With the Feynman trick, the result becomes
particularly simple in the chiral limit $q^2=m_\pi^2 \to 0$, yielding 
\begin{eqnarray}\
\widetilde \Psi_P(\alpha,x^2) &=& M \left[ -2 (2 \alpha-1) \partial_{x^2} 
+ \partial^\mu \partial_\mu + M^2\right] \partial_{M^2} \Delta (M,x),
\nonumber \\ \widetilde \Psi_A(\alpha, x^2) &=& M^2 \partial_{M^2} \Delta (M,x),
\nonumber \\ \widetilde \Psi_T(\alpha, x^2) &=& -2 \partial_{x^2} \partial_{M^2} \Delta
(M,x),
\end{eqnarray}
where we have introduced the free scalar propagator in the coordinate
space,
\begin{eqnarray}
\Delta(M,x)= \int \frac{d^4 p}{(2\pi)^4} \frac{e^{i p \cdot
    x}}{p^2-M^2} = \frac{M K_1 (M \sqrt{-x^2})}{4 \pi^2 \sqrt{-x^2}}.
\end{eqnarray}
From the previous formulas we find (for $m_\pi=0$) the relations
\begin{eqnarray}
\Psi_P(r)=2\Psi_T(r)&=&\frac{g_{\pi qq} N_c M}{\pi^2 r} 
  K_1(M r)\Big|_{\rm reg} \sim \frac{e^{-M r}}{r^{3/2}} \label{eq:NJL}
,  \\  \Psi_A(r) &=&\frac{g_{\pi qq} M N_c}{2
    \pi ^2}  K_0(M
  r)\Big|_{\rm reg} \sim \frac{e^{-M r}}{r^{1/2}},
 \nonumber 
\end{eqnarray}
where $K_0$ and $K_1$ are the modified Bessel functions and ``reg''
means a regulator. The asymptotic behavior at $r \to \infty$ is
independent of the regulator and implies a longer tail in the $A$
channel than in the $P$ and $T$ channels. The exponential fall-off of
the pion wave functions is consistent, up to a power in r, with the
probabilistic estimate in Eq.~(\ref{eq:prob-inv}) with $M_V = 2 M$,
$M$ being the {\it constituent} mass.  At short distances as $\Delta (x^2,M^2) \sim 1/x^2$ the results
are divergent demanding regularization.
 
The NJL model with the (twice subtracted) Pauli-Villars (PV)
regularization applied to an observable $A$ amounts to the replacement $M^2
\to M^2+\Lambda^2$, followed by the subtraction
\begin{eqnarray}
A|_{\rm reg}=A(\Lambda^2=0)-A(\Lambda^2)+\Lambda^2 \frac{dA(\Lambda^2)}{d\Lambda^2}. \label{PVpr}
\end{eqnarray}
Results of the calculation, shown already
in~\cite{Broniowski:2009dt}, are given in
Fig.~\ref{fig:njl-wf}. As we see, the agreement is excellent.  At the
farthest lattice point, $r=0.6~{\rm fm}$, the ratio $r \, \Psi_A(r)/
\Psi_P(r)$ approaches a constant.

The quantitative agreement of the NJL model with the data is not
trivial. For instance, in the Spectral Quark
Model (SQM)~\cite{RuizArriola:2003bs} only qualitative matching is achieved.
In this approach the regularization is introduced by replacing the
mass $M$ by the ``spectral'' mass $\omega$, using $g_{\pi qq}=
\omega/f_\pi$ and integrating over $\omega$ with a suitable weight
$\rho(\omega)$, which depends on the vector and scalar meson masses
($M_V$ and $M_S$), and also on a specified contour in the complex $\omega$-plane. That way, for instance, VMD of the pion form factor can be
built in with $M_V^2 = 24\pi^2 f_\pi^2/N_c$. For
$m_\pi=0$ we get the results 
\begin{eqnarray} 
\frac{\Psi_P (r)}{\Psi_P(0)} = \frac{\Psi_T (r)}{\Psi_T(0)} 
= e^{-M_S r /2},   \qquad 
\frac{\Psi_A (r)}{\Psi_A(0)} = e^{-M_V r /2} \left(1 + \frac{M_V r}{2}\right) .
\label{eq:SQM}
\end{eqnarray} 
Again, $\Psi_A$ is more extended than $\Psi_P$ and $\Psi_T$, due to
the presence of an extra power in $r$.
 In the case of SQM we would get good fits of the lattice
data for $M_V/2=505(30)$, $520(20)$, $530(14)$~MeV for the subsequent
values of $m_\pi= 345$, $475$, $575$~MeV. A
simple quadratic extrapolation in $m_\pi$ to the chiral limit yields $M_V/2=493(20)~{\rm
  MeV}$, a too high value as compared to 
$M_V=m_\rho=770~{\rm MeV}$.

\section{Holographic wave functions vs lattice}

\begin{figure}
\begin{center}
\epsfig{figure=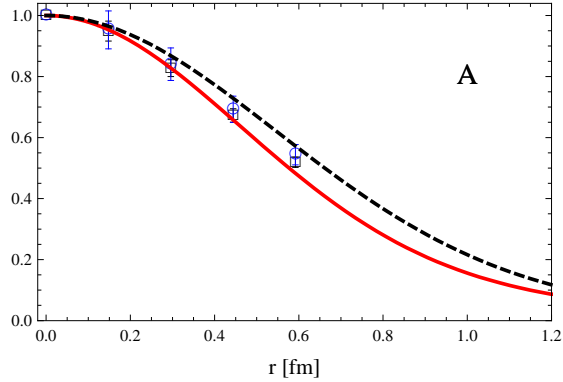,height=5cm,angle=0} 
\end{center}
\vspace{-6mm}
\caption{Axial equal-time holographic wave functions, normalized to
  unity at the origin, for the quark masses $m=330$~MeV (dashed black) and $m=4$~MeV (solid red), 
  obtained via
  the transversity relations and compared to the lattice calculation
  with $m_\pi=575$~MeV (boxes) and $345$~MeV
  (circles)~\cite{Broniowski:2009dt}.}
\label{fig:holo-wf}
\end{figure}

The transversity relations (\ref{eq:transv}) can be used to
deduce the ET wave functions from the LC wave functions.  As we have
shown, ET wave functions can be computed on Euclidean lattices upon a
suitable coarse graining of the gauge link operator. As an example, we
take here the holographic wave functions inspired by the AdS/CFT
correspondence and recently brought in connection with the LC wave
functions~\cite{deTeramond:2008ht} as a first approximation to QCD
(for a review see e.g. \cite{Brodsky:2010px}). The basic idea was to
relate the QCD-LC Hamiltonian by using the scaling relation between
the two-dimensional vectors ${\bf \zeta} = \sqrt{x(1-x)}\, {\bf b}$
(valid for massless quarks) and {\it assume} that the interaction also
depends on this scaling variable. Further elaborated models with
finite quark masses are introduced, assuming the replacement ${\bf
  k}_\perp^2 \to {\bf k}_\perp^2 + m^2$
~\cite{Vega:2009zb}. However, by doing so it is not obvious whether
$m$ corresponds to the current or constituent mass of the quark.  For
that reason, masses $m=4~{\rm MeV}$ and $m=330~{\rm MeV}$ are
explored. A full discussion of models is carried out
in~\cite{Afonin:2010fr}, where the soft-wall with a positive dilaton
background seems phenomenologically preferred.
Actually, besides the good quality of the mass spectrum, $M_{nLS}^2 =
2\pi \sigma (n+L+S/2)$ with $ 4 \kappa^2 = 2 \pi \sigma $ 
, the pion arises as a massless mode (corresponding
to $n=L=S=0$). Unlike the chiral models, this is not linked to
the spontaneous breaking of the chiral symmetry, apparently not
manifest in the light-cone dynamics.  Thus, we identify the BS axial
component with the holographic wave
function~\cite{deTeramond:2008ht,Vega:2009zb}
\begin{eqnarray}
\Psi_A^{\rm LC} ({\bf b}_\perp, x) = \frac{A\kappa}{\sqrt{\pi}}\sqrt{x (1 - x)}
\exp\left[ -\frac{\kappa^2}2 x (1 - x){\bf b}_\perp^2-\frac{m^2}{2 \kappa^2
    x(1-x)} \right].
\label{eq:psi-holo}
\end{eqnarray}
Calculations can be undertaken analytically for massless quarks.
Following~\cite{Vega:2009zb} we fix  the pion weak decay,
  $\pi^+ \to \mu^+ \bar \nu $, and the neutral pion decay, $\pi^0 \to
  2 \gamma$, from the conditions 
$$
\frac{1}{2 \sqrt{\pi}}\int_0^1 dx\, \Psi_A^{\rm LC} ({\bf b}_\perp , x)|_{{\bf b}_\perp=0_\perp} =
\frac{f_\pi}{2 \sqrt{3}} = \frac{A \kappa}{16} \, \, , \qquad 
2 \sqrt{\pi}\int_0^1 dx d^2 {\bf b}_\perp \, \Psi_A^{\rm LC} ({\bf b}_\perp , x) =
\frac{\sqrt{3}}{f_\pi} = \frac{4 A \pi^{2}}{\kappa} 
$$ 
respectively, 
with $f_\pi= 92.4~{\rm MeV}$. 
This yields $A^2 = 2
/\pi^2 = 0.2$ and $\kappa= 4 \pi \sqrt{2/3} f_\pi = 950~{\rm
  MeV}$.
Using the transversity relation the ET (holographic) wave
function reads (in the massless case, $m=0$)
\begin{eqnarray}
\frac{\Psi^{\rm ET}_A (r)}{\Psi^{\rm ET}_A (0)} = 
 e^{-2 f_\pi^2 \pi^2 r^2/3} \left[ 
I_0 (2 f_\pi^2 \pi^2 r^2/3)
-I_1 (2 f_\pi^2 \pi^2 r^2/3)
\right] = 1 - \pi^2 f_\pi^2 r^2 + {\cal O} (r^4), 
\end{eqnarray}
where $I_n(z)$ are the modified Bessel functions. Note that upon the
use of the relation $M_V^2 = 24 \pi^2 f_\pi^2/N_c$ the small-$r$
expansion reproduces the SQM result, Eq.~(\ref{eq:SQM}), exactly.  The
asymptotic behavior at large distances has the form
\begin{eqnarray}
\frac{\Psi^{\rm ET}_A (r)}{\Psi^{\rm ET}_A (0)} = 
\frac{3 \sqrt{3}}{8 f^3 \pi ^{7/2}}\left( \frac{m}{r^2} + \frac{1}{r^3} + {\cal O} (r^{-5}) \right )e^{-m r}  \,  ,
\end{eqnarray}
which displays the exponential fall-off, similarly to the chiral quark
models, Eq.~(\ref{eq:NJL}) for NJL and Eq.~(\ref{eq:SQM}) for SQM,
however the powers of $r$ are different, exhibiting different dynamics
in the models. The parameters $A$ and $\kappa$ for $m=4~{\rm MeV}$ and
$m=330~{\rm MeV}$ in Eq.~(\ref{eq:psi-holo}) are taken as in
\cite{Vega:2009zb}).  As we can see from Fig.~\ref{fig:holo-wf}, the
agreement is quite good and the lattice data hardly allow to
discriminate between the mass values except at $r=0.6~{\rm fm}$.  Of
course, it would be very useful to pin-down the correct long distance
behavior from the lattice above $0.6~{\rm fm}$. The calculation of the
other (higher-twist) components within the holographic approach would
also be highly desirable.

Our numerical study shows that the lattice data in
Fig.~\ref{fig:holo-wf} can be best fitted with $m \sim 100$~MeV,
similar to the current quark mass used in the NJL calculation at high
values of $m_\pi$. In particular, for $m_\pi=575$ and $345$~MeV we
have used $m=140$ and $51$~MeV, respectively, in rough agreement with
the Gell-Mann--Oakes--Renner relation $-m \langle \bar q q \rangle =
f_\pi^2 m_\pi^2$. Thus, the interpretation of $m$ in the holographic
models as the {\em current} quark mass seems consistent with this
comparison.

\section{Conclusions}

The presented calculations show that the Euclidean lattices can be
successfully used to coarse grain the wave function over
short-distance scales, where the gluon degrees of freedom are
integrated out. A direct comparison to wave functions from various
hadronic models not only becomes meaningful, but in some cases very
successful.  A quite unexpected result concerns the utility
of our calculations to determine the transversity information (the
dependence on the transverse coordinates), relevant for the light-cone
physics; the infinite-momentum frame pion wave functions, integrated
over the Bjorken-$x$, coincide in the impact-parameter space with the
equal-time rest-frame wave functions. This relation provides a way of
checking the wave functions for models genuinely formulated in the LC
variables. As an example, we have carried out this analysis for
holographic models.

We thank Sasa Prelovsek and  Luka Santelj for their collaboration in 
the lattice calculation~\cite{Broniowski:2009dt}. 

\bibliographystyle{JHEP-2}
 
\providecommand{\href}[2]{#2}\begingroup\raggedright\endgroup

\end{document}